\documentclass[manuscript]{aastex}

\usepackage{natbib}
\usepackage{graphicx}
\usepackage{epstopdf}
\usepackage{multirow}

\newcommand{\Swift}{{\em Swift }}

\shorttitle{New Derivation of Jet Opening Angles}
\shortauthors{Goldstein et al.}

\begin{document}

\title{A New Derivation of GRB Jet Opening Angles from the Prompt Gamma-Ray Emission}

\author{Adam Goldstein\altaffilmark{1}, 
Robert~D.~Preece\altaffilmark{1}, 
Michael~S.~Briggs\altaffilmark{1}, 
Alexander J.~van der Horst\altaffilmark{2}, 
Sheila McBreen\altaffilmark{3}, 
Chryssa Kouveliotou\altaffilmark{4}, 
Valerie Connaughton\altaffilmark{1}, 
William S.~Paciesas\altaffilmark{1}, 
Charles A.~Meegan\altaffilmark{2}, 
P.~N. Bhat\altaffilmark{1}, 
Elisabetta Bissaldi\altaffilmark{5}, 
J.~Michael Burgess\altaffilmark{1}, 
Vandiver Chaplin\altaffilmark{1}, 
Roland Diehl\altaffilmark{5}, 
Gerald~J.~Fishman\altaffilmark{4}, 
Gerard Fitzpatrick\altaffilmark{3}, 
Suzanne Foley\altaffilmark{5}, 
Melissa Gibby\altaffilmark{6}, 
Misty Giles\altaffilmark{6}, 
Jochen Greiner\altaffilmark{5}, 
David Gruber\altaffilmark{5}, 
Sylvain Guiriec\altaffilmark{1}, 
Andreas von Kienlin\altaffilmark{5}, 
Marc Kippen\altaffilmark{7},  
Arne Rau\altaffilmark{5}, 
Dave Tierney\altaffilmark{3}, and 
Colleen Wilson-Hodge\altaffilmark{4}}

\altaffiltext{1}{University of Alabama in Huntsville, 320 Sparkman Drive, Huntsville, AL 35899, USA}
\altaffiltext{2}{Universities Space Research Association, 320 Sparkman Drive, Huntsville, AL 35899, USA}
\altaffiltext{3}{University College, Dublin, Belfield, Stillorgan Road, Dublin 4, Ireland}
\altaffiltext{4}{Space Science Office, VP62, NASA/Marshall Space Flight Center, Huntsville, AL 35812, USA}
\altaffiltext{5}{Max-Planck-Institut f$\rm \ddot{u}$r extraterrestrische Physik (Giessenbachstrasse 1, 85748 Garching, 
Germany)}
\altaffiltext{6}{Jacobs Technology}
\altaffiltext{7}{Los Alamos National Laboratory, PO Box 1663, Los Alamos, NM 87545, USA}

\begin{abstract}
The jet opening angle of gamma-ray bursts (GRBs) is an important parameter for determining the characteristics of the 
progenitor, and the information contained in the opening angle gives insight into the relativistic outflow and the total energy 
that is contained in the burst.  Unfortunately, a confident inference of the jet opening angle usually requires broadband 
measurement of the afterglow of the GRB, from the X-ray down to the radio and from minutes to days after the prompt 
gamma-ray emission, which may be difficult to obtain.  For this reason, very few of all detected GRBs have constrained jet 
angles.  We present an alternative approach to derive jet opening angles from the prompt emission of the GRB, given that 
the GRB has a measurable $E_{peak}$ and fluence, and which does not require any afterglow measurements.  We present 
the distribution of derived jet opening angles for the first two years of the Fermi Gamma-ray Burst Monitor (GBM) operation, 
and we compare a number of our derived opening angles to the reported opening angles using the traditional afterglow 
method.  We derive the collimation-corrected gamma-ray energy, $E_{\gamma}$, for GRBs with redshift and find that 
some of the GRBs in our sample are inconsistent with a proto-magnetar progenitor.  Finally, we show that the use of the 
derived jet opening angles results in a tighter correlation between the rest-frame $E_{peak}$ and $E_{\gamma}$ than has 
previously been presented, which places long GRBs and short GRBs onto one empirical power law.
\end{abstract}


\section{Introduction}
The Gamma-Ray Burst Monitor (GBM) onboard the Fermi Gamma-Ray Space Telescope has detected over 500 GRBs in
its first 2 years of operation.  A forthcoming catalog \citep{GoldsteinGBM} contains time-integrated and time-resolved 
spectra for nearly all bursts during this time frame.  With 12 sodium iodide (NaI) detectors and two bismuth germanate 
(BGO) detectors, GBM covers a wide energy band from 8 keV up to 40 MeV with roughly 2000 square centimeters of total 
detector surface area \citep{Meegan}.  This energy range effectively samples the prompt emission of GRBs and allows for 
rapid all-sky triggering and monitoring.  To gather information about the GRB afterglow properties, redshift, and jet opening 
angle, other instruments are required.  Since GBM can only localize a burst to 4 degree accuracy including systematic 
uncertainties \citep{Briggs}, a simultaneous detection with \Swift is usually required to derive a precise location for follow-up 
observations.  \Swift comprises a Burst Alert Telescope (BAT), an X-Ray Telescope (XRT) and an Ultraviolet-Optical 
Telescope (UVOT) \citep{Barthelmy}.  The \Swift prompt energy coverage extends from 20 -- 150 keV, which does not allow 
for a comprehensive study of the higher energy prompt emission of GRBs which normally peaks at a few hundred keV.  
Therefore, it is obvious that the synergy between GBM and \Swift results in a better understanding of this phenomenon.  
However, since there are relatively few GRBs that have been simultaneously detected by \Swift and GBM, we are motivated 
to extend the ability to determine intrinsic properties of GRBs, such as jet opening angles and energetics, to the GBM 
observations of the prompt emission alone, using \Swift afterglow studies for calibration.  

Current GRB theories assume that the explosions are collimated rather than isotropic, because otherwise the extreme 
energy outflow in gamma-rays during the prompt emission would eliminate many of the viable stellar mass progenitor 
models \citep{Rhoads, Sari}.  In fact, \citet{Rhoads} and \citet{Sari} proposed observable, achromatic `jet' breaks in the 
broadband afterglow light curves of GRBs to distinguish the jet collimation opening angle before such breaks were 
discovered \citep{Granot}.  It is now widely accepted that GRBs are collimated, yet few well-constrained jet angles have 
been unambiguously  identified.  Further, nearly all estimated jet opening angles have been for the long soft class of GRBs, 
while there are no constrained estimates for the short hard class \citep{Kouveliotou}, most likely because short GRB 
afterglows are fainter and are thus less likely to be monitored long enough to detect a break \citep{Gehrels, Kann}.  In 
addition, precise locations are required and time must be requested of various observatories over a broad spectral range to 
study the late-time afterglow from minutes to days after the prompt emission.  Several observational constraints and effects 
can hamper the identification of jet breaks such as gaps in temporal and spectral coverage and the presence of optical 
bumps in the light curve and X-ray flares.  Even if a jet break is detected, an assumption of the density of the circumburst 
medium is required \citep{Sari, Chevalier}, and so a certain amount of uncertainty is inherent in the calculation of the jet 
opening angle.

Once determined, jet opening angles can lead to an estimate of the total energy release in gamma-rays.  If the 
redshift and the prompt emission fluence of the GRB are also known, the collimation-corrected energy release at the 
source will provide the total energy budget of the GRB.  These results constrain progenitor models and provide an estimate 
of the bulk Lorentz factor of the ejecta \citep{Granot}.  In addition, a more reliable and robust study of cosmology would be 
possible if a large number of collimation-corrected energies were known \citep{Bloom}.  Such studies are currently 
incomplete due to small statistics.  For example, out of nearly 500 GRBs detected by GBM through July 2010, only 30 have 
observed redshifts from the simultaneous detection by \Swift, and of these 30 GRBs only 8 have inferred jet opening angles 
from afterglow studies.  The importance of GRBs to cosmology has recently become very clear with the detection of bursts 
out to a $z$ of 8.2 \citep{090423}, making these events among the farthest detected in the observable Universe.  Many 
authors have studied the spectral properties such as the peak luminosity;  isotropic energy release in gamma-rays, $E_
{iso}$; the peak energy of the GRB power density spectrum, $E_{peak}$ \citep{Mallozzi, Koshut, Lloyd, Amati, Bloom, 
Ghirlanda, Yonetoku, Firmani}.  The purpose of this paper is to show that we can provide an alternative to the afterglow 
lightcurve monitoring method by deriving jet opening angles from the prompt emission of GRBs.  We then utilize the 
opening angles to estimate the collimation-corrected energy release in gamma-rays, $E_{\gamma}$, and show that there is 
a tight correlation between $E_{peak}$ and $E_{\gamma}$. Section 2 describes the data sample and section 3 consists of 
our data analysis methods and results. Finally, we discuss the cosmological consequences and implications for progenitors 
of our study in section 4.

\section{Data Sample}
Two data samples were used in this study.  The `redshift' sample consists of 30 bursts from the first 2-year GBM 
catalog as well as 3 more bursts in the third year (GRB 100724A, 100814A, and 100816A).  All other 
bursts (without redshift) were chosen by selecting events from the entire GBM catalog from the first two years \citep
{GoldsteinGBM} according to the following data selection cuts.  First, the spectral data were retrieved from fits performed 
with the standard GBM spectral fitting program, RMfit.  Two models were investigated: the Band GRB function \citep{Band}, 
which comprises two power laws smoothly joined at a break energy that is unique to each burst; and a `Comptonized' 
model, which is a power law with a high-energy exponential cutoff.  Both models were parametrized with the $E_{peak}$ 
parameter, which is the energy at which peak power production is measured.  Both models well describe nearly 80\% of all 
GRB spectra, though the both functions are purely empirical and are not derived from physical quantities.  The fitting 
statistic used in RMfit is the Castor C-statistic, a modification of the Cash statistic \citep{Cash}.  For each burst, a model 
comparison was performed by calculating the change in likelihood between the two models and determining the 
corresponding chi-square distribution for 1 degree of freedom (as the Comptonized model has one more degree of 
freedom and is nested within the Band model).  We choose a change in likelihood of 6 units per one degree of freedom, 
which has a chance probability of about $.01$, as a threshold for a model to be preferred. The values of interest in our 
study are $E_{peak}$ and fluence, so a data cut was performed on the error in  $E_{peak}$ and the fluence.  Only those 
bursts that had an error of 40\% or less of the mean value of the quantity of interest were allowed into the sample.  This was 
done to ensure the integrity of our data sample and the following analysis.

\section{Data Analysis and Results}

\subsection{$\bf E_{peak}$ \& Fluence}
To begin, we investigate a new discriminator between two types of bursts (long and short), the $E_{peak}$/Fluence energy 
ratio \citep{Goldstein}, which was discovered using BATSE data.  This ratio provides a measure of spectral hardness similar 
to that found by \citet{Kouveliotou}, but it is independent of redshift in energy and is directly related to the luminosity  
distance.  Shown in Figure 1 is the distribution of 382 GRBs from the GBM spectral catalog \citep{GoldsteinGBM}.  Using 
the preliminary duration results from the upcoming GBM GRB Catalog \citep{Paciesas}, the figure strongly supports the 
original claim by \citet{Goldstein} and shows the distribution separated into long bursts and short bursts, and a 
correspondence has been found between the energy ratio and the observed duration estimate of the bursts.  

In an attempt to relate the rest-frame $E_{peak}$ with the total energy release in gamma-rays, \citet{Amati} discovered a 
correlation between the rest-frame $E_{peak}$ and $E_{iso}$, the bolometric energy release in gamma-rays assuming 
isotropic emission.  The correlation is highly susceptible to scatter and outliers, and may be a result of selection effects due 
to detector trigger and spectral criteria \citep{Lloyd-Ronning, FriedmanBloom,NakarPiran,  BandPreece}.  \citet{Ghirlanda} 
found a similar correlation using $E_{\gamma}$, the collimation-corrected energy release.  The so-called Ghirlanda 
relation contains less scatter and fewer outliers, but requires an additional piece of information, the jet opening angle.  Both 
of these correlations require the redshift of the GRB, yet previous papers \citep{NakarPiran, BandPreece} have devised a 
way to test the correlations with GRBs that have no observed redshift.  We adopt this test to show in Figure 2 the lower limits 
of the Amati and Ghirlanda relations in the $E_{peak}$--Fluence plane, and we find that very few bursts ($\sim18$\%) can 
follow the Amati relation, though all GBM bursts may be valid for the Ghirlanda relation.  More importantly, we plot the long 
and short bursts separately for GBM and we find that most long bursts are clustered between the Amati and Ghirlanda 
lower limits, while most of the short bursts are linearly dispersed along the Ghirlanda lower limit.  We assume a beaming 
factor of unity for the Ghirlanda relation in the Figure 2, indicating that the opening jet angle is 90 degrees, which is 
consistent with previous findings from BATSE \citep{Goldstein}.  Interestingly, if we decrease the beaming factor (and thus 
the jet opening angle), the Ghirlanda lower limit moves towards the bulk of long bursts, and eventually all short bursts will 
violate the Ghirlanda lower limit.  This gives support to the findings that long bursts have a dispersion of jet angles from 
small angles on the order of a degree up to 50 degrees \citep{Frail, Nakar, Panaitescu} and short bursts have larger 
average jet opening angles of 40-90 degrees, as is supported theoretically by \citet{Livio} and observationally by \citet
{Watson}.

\subsection{Jet Opening Angles}
As indicated, the combination of the $E_{peak}$--Fluence plane and the Ghirlanda lower limit admit derived jet opening 
angles for GRBs, since the Ghirlanda lower limit appears to describe a true cutoff in the $E_{peak}$--Fluence plane that is 
consistent among multiple instruments.  We can solve the lower  limit from Ghirlanda's best fit equation \citep{Ghirlanda}, 
which was calibrated with GRBs that had well-constrained jet opening angles, in terms of the beaming fraction, observed 
$E_{peak}$, and fluence (see \citet{BandPreece, Goldstein}).  Using this tool, we can derive jet opening angles for a large 
number of bursts with constrained $E_{peak}$  values and fluences.  In Table 1 we show the reported jet opening angles 
for bursts from afterglow studies in our redshift sample and compare them to our derived jet opening angles.  There was no 
data selection cut based on parameter error due to the small sample size, therefore Table 1 contains a few angles with 
large error bars.  In a number of cases, two different angles were inferred for a single burst, sometimes in conflict with each 
other.  It should be noted that \citet{Cenko} used radio observations to constrain the physical parameters in their sample, 
while radio measurements were not used by \citet{McBreen} or \citet{Rau}.  Using only the four bursts with well-constrained 
reported jet angles, the Spearman's rank correlation coefficient for the reported angles and the derived angles is 0.8 with a 
probability of 0.2 that the two quantities are not correlated given the null hypothesis.  Encouraged by the good agreement 
(within errors) between our estimates and the reported values of jet opening angles, we can now determine the jet angle for 
bursts without a measured redshift.  In Figure 3 we show the jet opening angles for 382 GRBs from the GBM spectral 
catalog that have $E_{peak}$  and fluence errors less than 40\% of the mean value, resulting in an error in jet opening 
angle of no more than 40\% for 85\% of our calculated values.  The distribution for long GRBs is extremely similar 
to the inferred range of opening angles from afterglow studies (2-50 degrees), while the short GRBs cluster closer to 90 
degrees.

\subsection{Energetics \& Correlations}
We now proceed to explore the actual Amati and Ghirlanda relations with our data.  In Figure 4 we show the rest-frame $E_
{peak}$  versus the estimated isotropic energy release, $E_{iso}$ for the GBM redshift sample.  This is a direct test of the 
$E_{peak}$--$E_{iso}$  relations with bursts of known redshift, however we can use the derived upper limits of the relations 
to help constrain the correlations.  From Figure 4, we find the long bursts are loosely dispersed along the Amati relation 
\citep{Amati}, yet a number of long bursts are situated several sigma above the Amati upper limit (dotted line), which implies 
that if the bursts follow the Amati relation, their redshifts would be imaginary numbers. Interestingly, the Ghirlanda relation 
\citep{Ghirlanda}well describes the linear dispersion of short hard GRBs if a jet opening angle of 90 degrees is assumed.  
This lends credence to the idea that short bursts are near isotropic bursters \citep{Livio, Watson}.

Using our derived jet opening angles, we can show that there is a correlation between the rest-frame $E_{peak}$ and 
the total energy release in gamma-rays after correcting for collimation, $E_\gamma$, which is tantamount to a 
luminosity-distance relation.  In previous attempts at such a relation, \citet{Amati} studied $E_{iso}$, which would seem to 
define a total energy budget for the explosion, but obtained a much looser correlation, and was unable to apply the relation 
to short GRBs.  \citet{Ghirlanda}, however, studied inferred $E_\gamma$ obtained from jet opening angles deduced from 
broadband afterglow observations of a few GRBs and found a tighter relation than \citet{Amati}; they were still unable to fit 
short GRBs into the relation.  Figure 5 shows the distribution of $E_{\gamma}$ from the derived jet opening angles, which is 
wider than previous estimates \citep{Frail, Panaitescu} at 5 orders of magnitude and appears to peak near $1 \times 10^
{51}$ erg, although this may be due to a larger sample size compared to those previous studies.  For a few bursts in the 
redshift sample, $E_{\gamma}$  surpasses the total energy limit for a proto-magnetar, which is a few$\times 10^{52}$ erg 
\citep{Metzger} and a number of other bursts would require a very high efficiency to have viable magnetar progenitors, 
therefore according to our results, it is doubtful that a number of GRBs in our redshift sample originate from magnetar 
progenitors.  The energy budget for neutron star mergers is in the range of a few$\times 10^{53}$ erg \citep{Woosley}, while 
theoretically collapsars could emit up to $10^{54}$ erg \citep{Woosley, Paczynski} at extremely high efficiency, which is 
compatible with our results.

Using the derived $E_{\gamma}$ obtained from the clear cutoff in the $E_{peak}$--Fluence plane, Figure 6 shows the new 
relation  and provides a much tighter correlation than either of the two previous relations. It also shows that short GRBs are 
clearly not outliers after correction for the much wider beaming angle.  We present the best fit power law as
\begin{equation}
	E^{rest}_{peak}=(589 \pm 18 \; \rm keV) \Biggl( \frac{E_{\gamma}}{10^{51} \; \rm erg} \Biggr)^{0.49 \pm 0.01}
\end{equation}
where the empirical power law was fitted to both coordinates.  The power law normalization is of the same order as that 
found by \citet{Ghirlanda} (480 keV), but the power law index is very similar to that found by \citet{Amati} (0.52).  The 
Spearman's rank correlation coefficient for $E_{peak}$ and $E_{\gamma}$ is 0.91 with a probability of $3 \times 10^{-13}$ 
that the two quantities are not correlated given the null hypothesis.  The calculated scatter around the best fit power law is 
about 0.1 dex for bursts with redshift  $\sim 0.5 \le z \le 8.2$ and covering over 4 orders of magnitude in energy.

\section{Conclusions}
We have confirmed the $E_{peak}$/Fluence energy ratio results for GBM bursts, and have shown how they can be related 
to two different classes of GRBs.  Most likely the larger of the two distributions belongs primarily to the long class of bursts 
typically associated with low-metallicity core-collapse supernovae, while the smaller mode belongs to the short class 
which is believed to be associated with neutron star-neutron star and neutron star-black hole mergers. The distribution of 
the energy ratio provides information for the determination of the rest-frame energetics, as can be seen in more detail when 
plotted in the $E_{peak}$--Fluence plane.  The hard spectral cutoff in this plane is apparent for all BATSE and GBM bursts, 
despite the fact that these were detected with two different instruments with different sensitivities and different band passes. 
From this we can infer the cutoff is most likely not detector dependent, and using only physical observables, we can derive 
one of the most important parameters of GRBs, the jet opening angle, and given an observed redshift, calculate $E_
{\gamma}$.  

Previously, the jet opening angle was only inferred for a handful of bursts for which X-ray,  optical, and radio measurements 
of the afterglow were available \citep{Sari}.  Even if these measurements were available, very few bursts have discernible 
jet breaks that denote the moment at which the relativistic ejecta slow down to the point that the observed relativistic 
beaming angle is the same as the actual beaming angle of the outflow \citep{Granot}.  \citet{Ghirlanda05} derived jet 
opening angles without afterglow measurements, although their derivation required the bursts to be described by the 
Ghirlanda relation and relied on the lag-luminosity relationship \citep{Norris} to derive pseudo-redshifts, yielding an 
ensemble distribution of jet opening angles.  From this, they postulated that bursts with softer $E_{peak}$ have larger 
opening angles than bursts with harder $E_{peak}$.  By traditional classification of GRBs \citep{Kouveliotou}, the 
interpretation is that short hard GRBs would have smaller opening angles than long soft GRBs.  This is, in general, 
contradictory to our findings.  To contrast, our results show that we can reproduce jet opening angles for individual GRBs 
that can be spectrally analyzed with prompt emission alone and does not require the estimation of pseudo-redshifts.  Our 
resulting distribution of opening angles also agrees with a current theoretical model that the beamed outflow at the 
rotational poles of the progenitor is produced by the rotational angular momentum, as well as the configuration of the 
magnetic field \citep{Livio}.  In this model, the degree of collimation is related to the ratio of radius of the compact object to 
the radius of the accretion disk.  In the case of a collapsar, the radius of the central object is much smaller than the radius of 
the disk, resulting in a tightly collimated beam, while a merger model results in much less collimation since the radius of the 
accretion disk is on the order of the radius of the central object.  Due to the much larger opening angles, observations of jet 
breaks for most short hard GRBs as well as some from the long soft class are unlikely since the estimated jet break times 
will be on the order of several months \citep{Sari}. 

Applying the jet opening angle to the redshift sample, we can estimate the entire energy outflow in gamma-rays.  From our 
results, we can rule out a proto-magnetar progenitor for a few long GRBs (080916C, 080810, 090323, 090519, and 
090902B) as well as a short GRB (090510).   When we compare this energy budget for each GRB to the $E_{peak}$, we 
find a clear correlation between the energy of peak power production and the total energy output in gamma-rays.  One 
major distinction of our relation shows that both classes of GRBs can be described by the same power law fit, which has not 
been previously shown. The uncertainties of our  calculations include geometric effects, as GRBs are likely seen slightly off-
axis from the center of the jet, and this error propagates through to the calculation of $E_\gamma$. These uncertainties will 
likely be small for long bursts with small opening angles, due to the small possible displacement of the viewing angle 
relative to the jet opening angle.  This relationship may potentially be exploited to extrapolate a rough redshift distribution 
for GRBs without redshift estimates, although a larger sample of GRBs with measured redshift is desired to confirm the 
existence of such a relation between $E_{peak}$ and $E_{\gamma}$.

\section{Acknowledgements}
A.G. acknowledges the support of the Graduate Student Researchers Program funded by NASA as well as the support and recommendations of the GBM Science \& Support Team.


\clearpage

\begin{figure}[ht]
\includegraphics[scale=.35, angle=270]{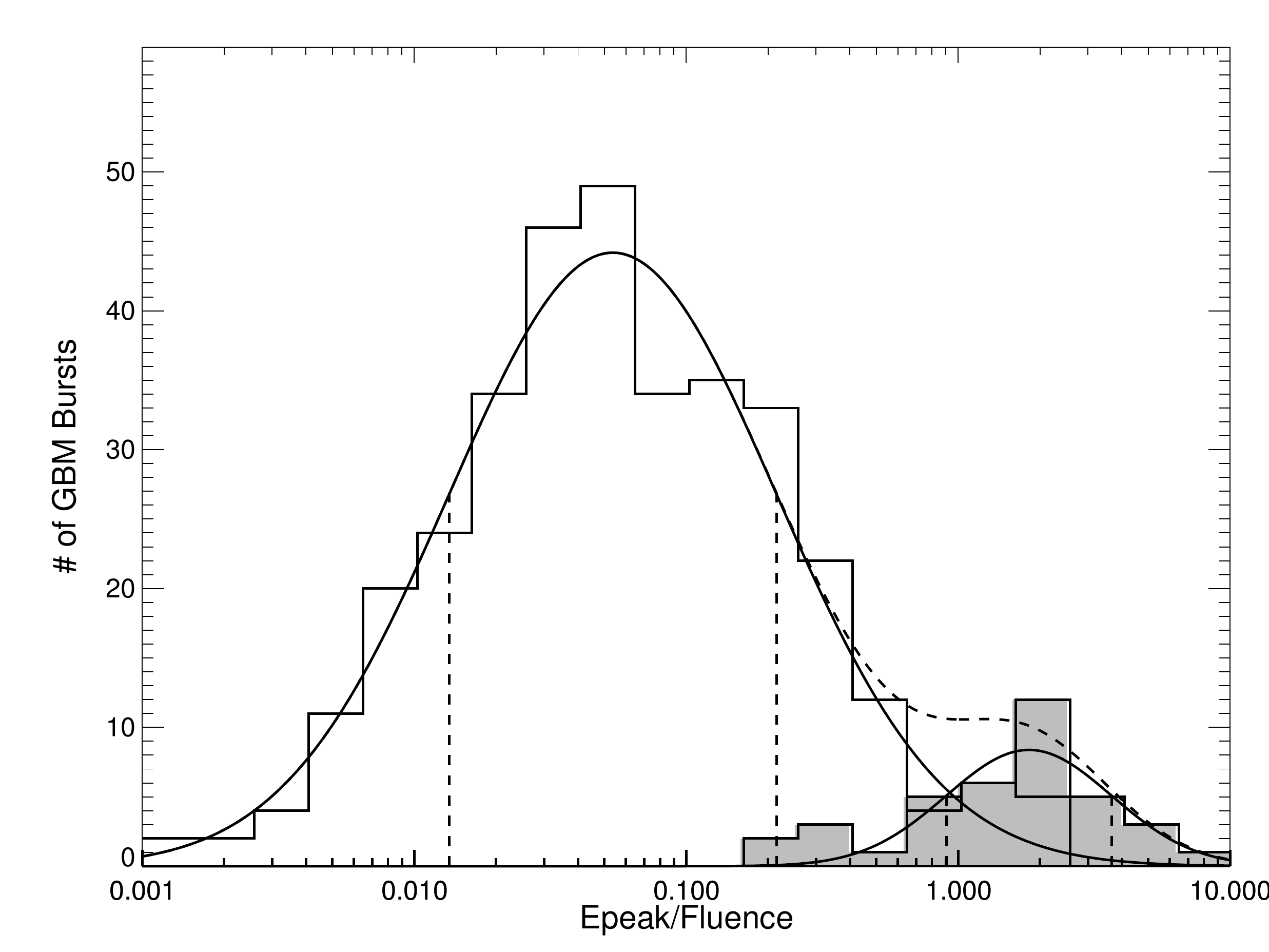}
\caption{ \label{fig1}The $E_{peak}$/Fluence energy ratio for 344 long GRBs (white) and 38 short GRBs (gray).  A fit was 
performed on the entire distribution and the $\chi^2$ goodness-of-fit for the two lognormal functions is 12.2 for 14 degrees 
of freedom.  Fitting a single lognormal function to the distribution results in a goodness-of-fit of 27.8 for 17 degrees of 
freedom.  The change in $\chi^2$ per degree of freedom results in a chance probability of $1\times 10^{-3}$ that the two
 lognormals are not preferred.  The vertical dotted lines denote the 1-$\sigma$ standard deviation of each lognormal.}
\end{figure}

\begin{figure}[ht]
\includegraphics[scale=.40, angle=270]{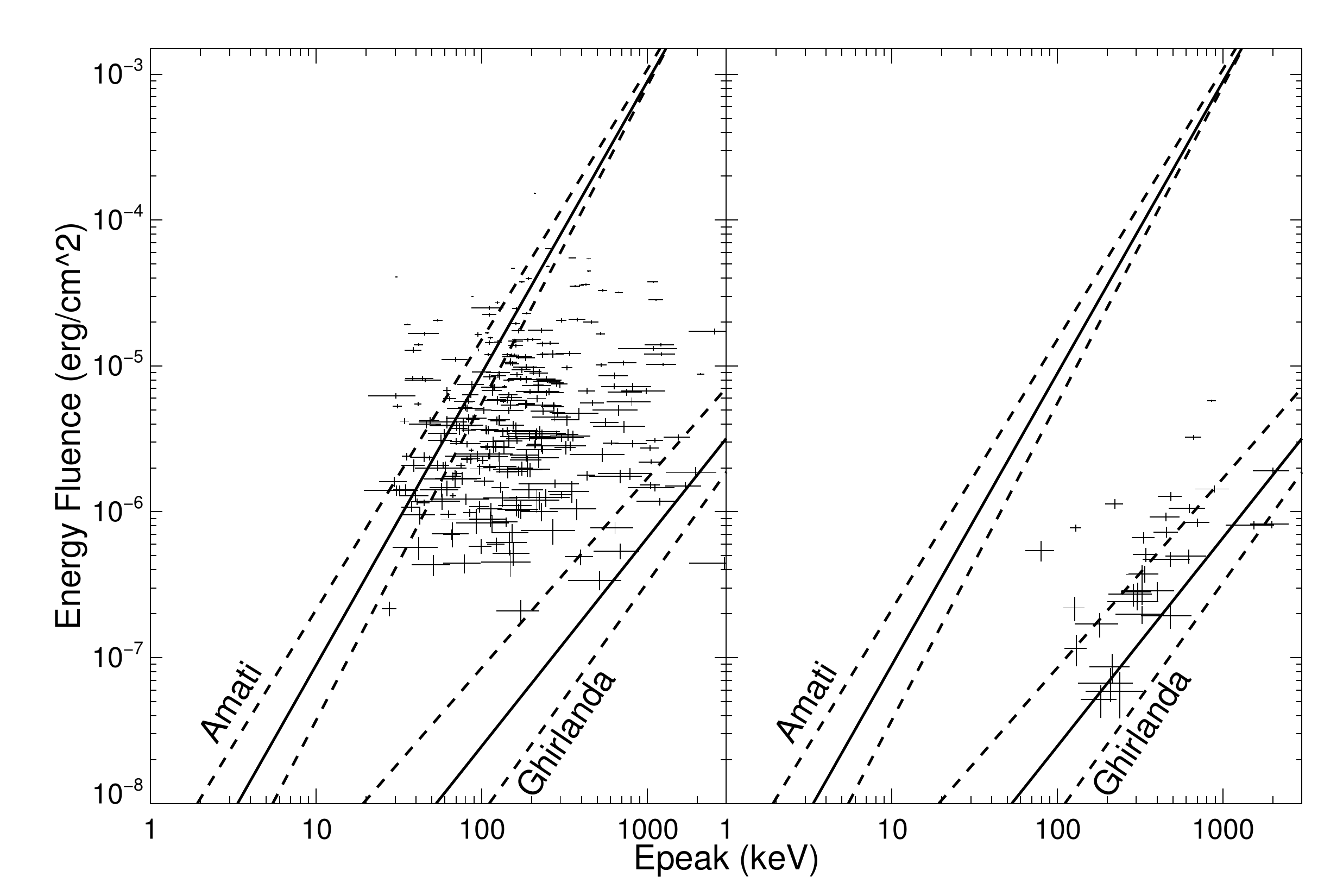}
\caption{ \label{fig2}Plot of 382 GBM bursts in the $E_{peak}$-Fluence plane.  The left side shows the 344 long GRBs and 
the right shows the 38 short.  The plotted lines are the Amati and Ghirlanda upper limits, which in this plane become lower 
limits.  From this we see that very few GRBs follow the Amati relation, since a majority of the bursts fall below the lower limit.  
The Ghirlanda lower limit, however, appears to be near a true lower bound.  Therefore, the Ghirlanda limit can be shown to 
admit a large spread of mostly small jet opening angles for long bursts and mostly large opening angles for short bursts. }
\end{figure}

\begin{figure}[ht]
\includegraphics[scale=.35, angle=270]{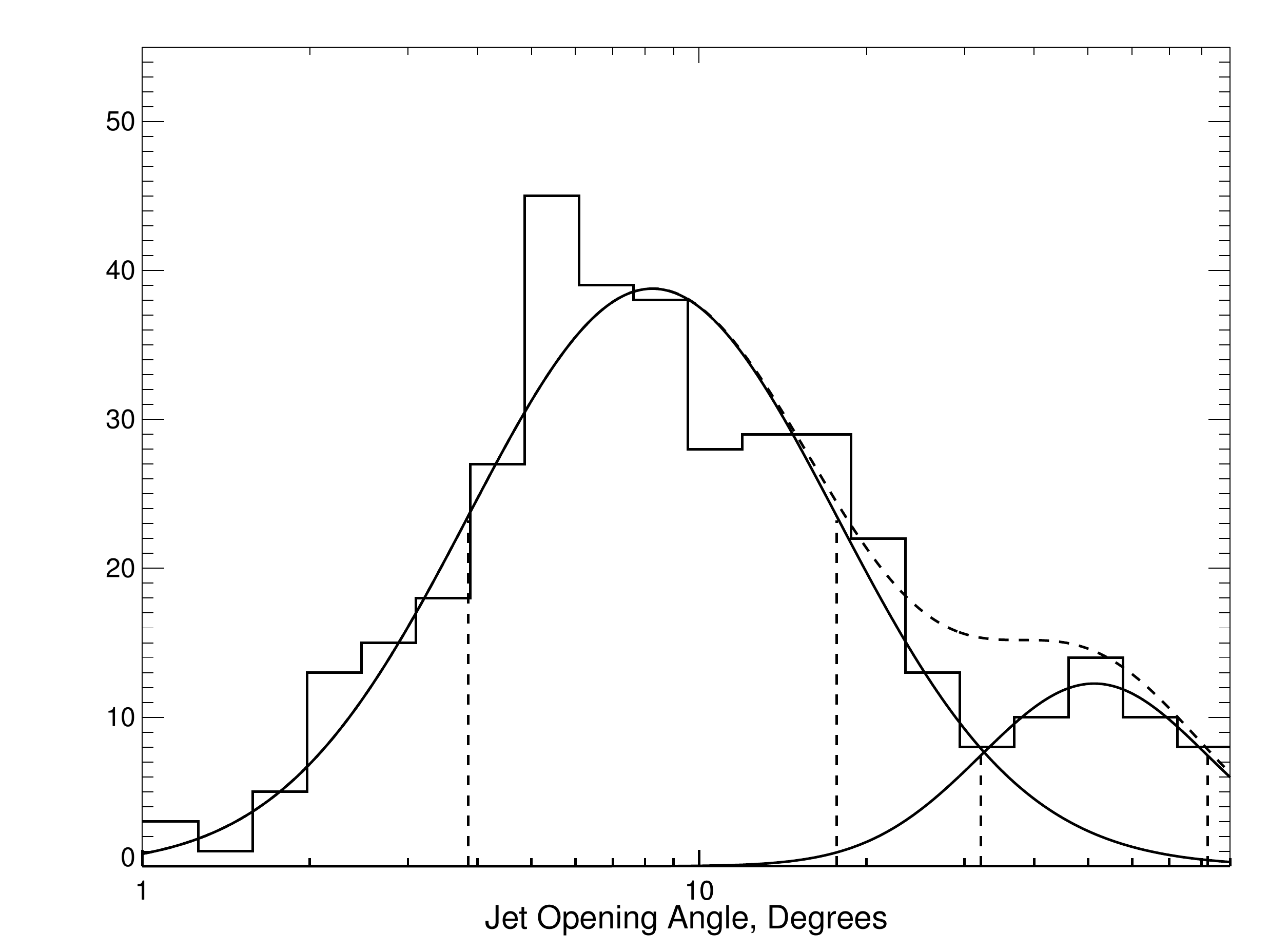}
\caption{ \label{fig3} Distribution of derived jet opening angles for 382 GBM GRBs.  The relative propagated errors for this 
plot do not exceed 0.40 for 85\% of our sample, and 51\% of our sample do not exceed a relative error of 0.1.  The best fit 
lognormal functions have a$\chi^2$ goodness-of-fit of 12.9 for 14 degrees of freedom.  The mean value for the large 
distribution is $\sim8$ degrees and the mean for the smaller distribution $\sim51$ degrees.  The vertical dotted lines show 
the 1-$\sigma$ standard deviations of the respective distributions.}
\end{figure}

\begin{figure}[ht]
\includegraphics[scale=.40]{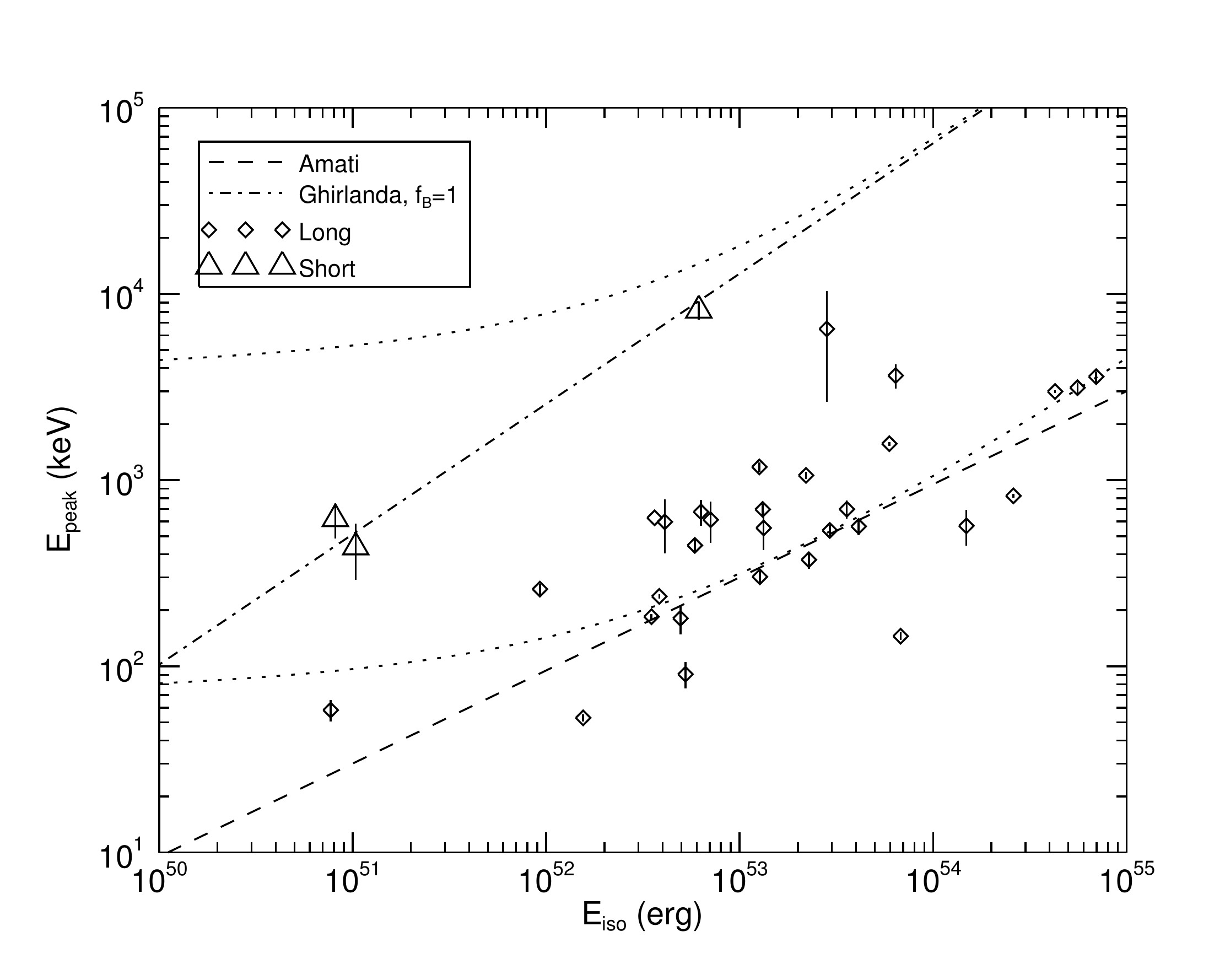}
\caption{ \label{fig4}Plot of the $E_{peak}$--$E_{iso}$ plane.  The dotted lines denote the upper limits for the two well-noted 
correlations.  Most long bursts cluster around the Amati line, but with a large dispersion.  The short bursts are situated along 
the the Ghirlanda line with corresponding jet opening angle of 
90 degrees.}
\end{figure}

\begin{figure}[ht]
\includegraphics[scale=.40]{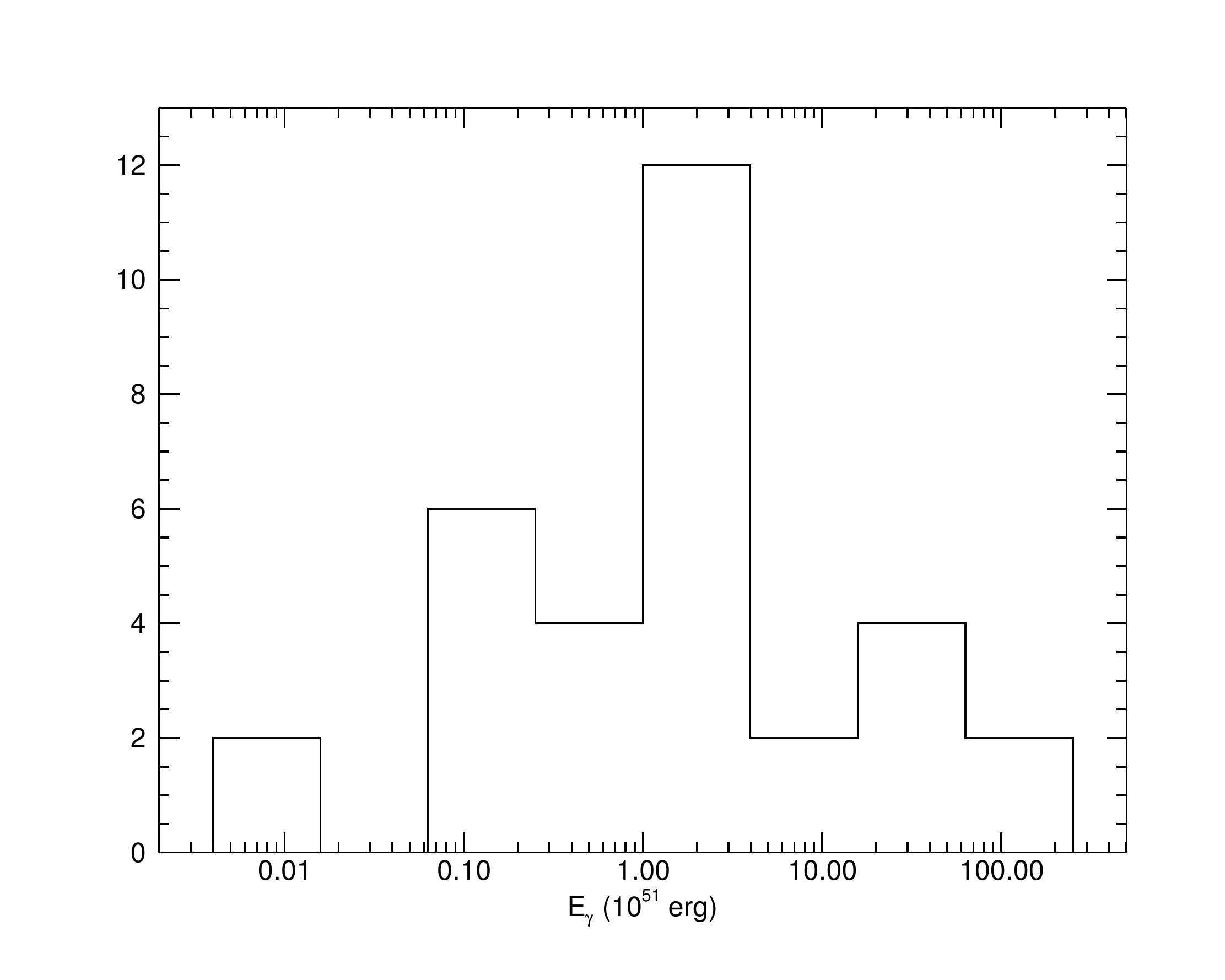}
\caption{ \label{fig5} Plot of $E_{\gamma}$ in units of $10^{51}$ erg from the redshift sample.  The distribution covers about 
5 orders of magnitude and peaks at $1 \times 10^{51}$ erg.  The bursts with $E_{\gamma}$ greater than $\sim3\times10^
{52}$ are inconsistent with the magnetar progenitor model}
\end{figure}

\begin{figure}[ht]
\includegraphics[scale=.40]{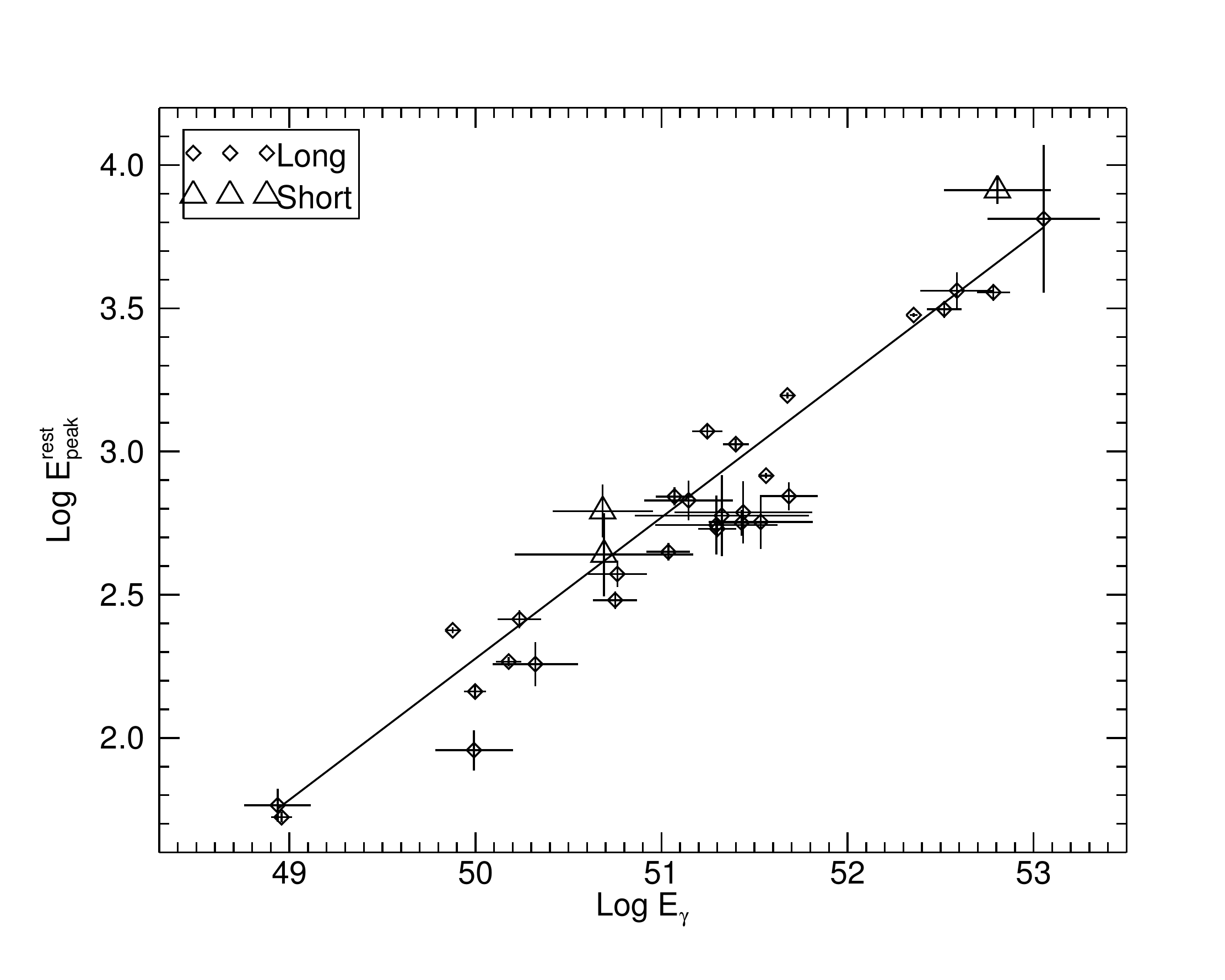}
\caption{ \label{fig6} Plot of the rest-frame $E_{peak}$ versus the total energy released in gamma-rays after correcting for 
collimation for the redshift sample.  The best fit value of the normalization of the power law is $589 \pm 18$ kev and the 
best fit power law index is 0.49 $\pm$ 0.01.  Note that the short GRBs follow the same correlation as long GRBs, which 
previous relations have been unable to show.  The relation covers over 4 orders of magnitude in energy and spans GRBs 
from $z$ of 0.5 up to 8.2. }
\end{figure}

\clearpage
\begin{deluxetable}{| c | c | c | c |}
\renewcommand{\thefootnote}{\alph{footnote}}
\tablecolumns{4}
\tablewidth{0pc}
\tablecaption{Comparison of Reported and Derived Jet Opening Angles}
\startdata
	\hline \bf GRB & \bf Reported Angle(s) & \bf Derived Angles & \bf Reference\\ 
	\hline 080810 & $> 4.0$  \tablenotemark{a} & $13.8 \pm 6.4$ & \citet{Page} \\ 
	\hline 080916C & $> 6.1$   \tablenotemark{b}& $5.2 \pm1.6$ & \citet{Greiner} \\ 
	\hline 081008 & $> 2.1$\tablenotemark{b} & $6.4 \pm 4.0$ & \citet{Yuan} \\ \hline
	\multirow{2}{*}{090323} & $< 2.1$ \tablenotemark{b} & \multirow{2}{*}{$4.3 \pm 1.3$} & \citet{McBreen} \\
					      & $2.6^{+0.6}_{-0.1}$ \tablenotemark{a} & & \citet{Cenko} \\ \hline
	\multirow{2}{*}{090328} & $< 5.5$ \tablenotemark{b} & \multirow{2}{*}{$6.6 \pm 11.9$} & \citet{McBreen} \\
					      & $5.2^{+1.4}_{-0.7}$ \tablenotemark{a} & & \citet{Cenko} \\
	\hline 090423 & $> 12.0$ \tablenotemark{b} & $11.0 \pm 7.0$ & \citet{Chandra} \\ \hline
	\multirow{2}{*}{090902B} & $> 6.4$ \tablenotemark{b} & \multirow{2}{*}{$4.1 \pm 0.6$} & \citet{McBreen} \\
					         & $3.4^{+0.4}_{-0.3}$ \tablenotemark{b }& & \citet{Cenko} \\ \hline
	\multirow{2}{*}{090926A} & $> 9.9$ \tablenotemark{b} & \multirow{2}{*}{$7.6 \pm 2.6$} & \citet{Rau} \\
					         & $7.0^{+3.0}_{-1.0}$ \tablenotemark{a} & & \citet{Cenko} \\
	\hline	
\enddata
\tablenotetext{a}{Wind Medium}
\tablenotetext{b}{ISM}
\end{deluxetable}

\end{document}